\documentclass[useAMS,usenatbib]{mn2e}
\usepackage{epsfig}
\usepackage{amsmath}

\newcommand{\be}{\begin{equation}}
\newcommand{\beq}{\begin{equation}}
\newcommand{\ba}{\begin{eqnarray}}
\newcommand{\ee}{\end{equation}}
\newcommand{\eeq}{\end{equation}}
\newcommand{\ea}{\end{eqnarray}}

\newcommand{\hs}{\hspace{1mm}}

\newcommand{\apj}{ApJ}
\newcommand{\aap}{A\&A}
\newcommand{\apjl}{ApJL}
\newcommand{\mnras}{MNRAS}
\newcommand{\aj}{AJ}
\newcommand{\apjs}{ApJS}
\newcommand{\nat}{{\it Nature}}
\newcommand{\araa}{ARA\&A}
\newcommand{\pasj}{PASJ}

\newcommand{\physrep}{Physics Reports}
\def\lsim{~\rlap{$<$}{\lower 1.0ex\hbox{$\sim$}}}

\def\gsim{~\rlap{$>$}{\lower 1.0ex\hbox{$\sim$}}}
\title[Ly$\alpha$ Radiation Pressure]{Ly$\alpha$ Driven Outflows Around Star Forming Galaxies}

\author[Mark Dijkstra \& Abraham Loeb]{Mark Dijkstra\thanks{E-mail:mdijkstr@cfa.harvard.edu} \& Abraham Loeb\thanks{E-mail:aloeb@cfa.harvard.edu}\\
Harvard-Smithsonian Center for Astrophysics, 60 Garden Street, Cambridge, MA 02138, USA}

\def\LaTeX{L\kern-.36em\raise.3ex\hbox{a}\kern-.15em
    T\kern-.1667em\lower.7ex\hbox{E}\kern-.125emX}

\begin{document}

\date{\today}
\pagerange{\pageref{firstpage}--\pageref{lastpage}} \pubyear{2007}

\maketitle

\label{firstpage}
\begin{abstract}
We present accurate Monte-Carlo calculations of Ly$\alpha$ radiation
pressure in a range of models which represent galaxies during various
epochs of our Universe. We show that the radiation force that Ly$\alpha$
photons exert on hydrogen gas in the neutral intergalactic medium (IGM),
that surrounds minihalos that host the first stars, may exceed gravity by
orders of magnitude and drive supersonic winds. Ly$\alpha$ radiation
pressure may also dominate over gravity in the neutral IGM that surrounds
the HII regions produced by the first galaxies. However, the radiation
force is likely too weak to result in supersonic outflows in this
case. Furthermore, we show that Ly$\alpha$ radiation pressure may drive
outflows in the interstellar medium of star forming galaxies that reach
hundreds of km s$^{-1}$. This mechanism could also operate at lower
redshifts $z\lsim 6$, and may have already been indirectly detected in the
spectral line shape of observed Ly$\alpha$ emission lines.  \end{abstract}

 
\section{Introduction}
\label{sec:intro}

HII regions around massive stars convert a significant fraction of the
total bolometric luminosity of young galaxies into Ly$\alpha$ line emission
\citep{PP67,S03}. This Ly$\alpha$ radiation can exert a large force on
surrounding neutral gas, as the Ly$\alpha$ transition has a
cross-section that is $\sim 7$ orders of magnitude
larger than the Thomson cross-section, when averaged over a frequency band as wide as the resonance frequency itself \citep[e.g.][]{Loeb01}. Not
surprisingly, the impact of Ly$\alpha$ radiation pressure on the formation
of galaxies has been discussed extensively
\citep[e.g][]{Cox85,EF86,Bithell90,Haenelt95,Oh02,McKee07}, but the
intricacies of Ly$\alpha$ radiative transfer in 3D complicated an accurate
numerical treatment of its dynamical effect on the gas. Nevertheless, an
approximate estimate can be obtained from simple energy considerations as
shown below.

Consider a self-gravitating gas cloud of total (baryons + dark matter) mass $M$ and radius $R$ that
contains a central Ly$\alpha$ source. The gravitational binding energy of
the baryons inside the cloud, $E_B\sim \Omega_bGM^2/(\Omega_m R)$, can be compared to the total energy in the
Ly$\alpha$ radiation field inside the cloud, $E_{\alpha}=L_{\alpha}\times
t_{\rm trap}$. Here, $L_{\alpha}$ is the Ly$\alpha$ luminosity of the
central source (in erg s$^{-1}$), and $t_{\rm trap}$ is the typical
trapping time of Ly$\alpha$ photons in the cloud owing to scattering on
hydrogen atoms. The Ly$\alpha$ radiation pressure would unbind the baryonic gas
from the cloud if $E_{\alpha}> E_B$, i.e. $L_{\alpha}>\Omega_bGM^2/(\Omega_mRt_{\rm trap})$
\citep[e.g.][]{Cox85,Bithell90,Oh02}. In this approach, $t_{\rm trap}$ is
one of the key parameters in setting the Ly$\alpha$ radiation
pressure. Calculations by \citet{Adams75} imply that $t_{\rm trap}\sim 15
t_{\rm light}$ for $3\lsim \log \tau_0 \lsim 5.5$, and $t_{\rm trap}\sim
15(\tau_0/10^{5.5})^{1/3} t_{\rm light}$ otherwise, for a static, uniform,
infinite slab of material \citep[also see Fig~1 of][]{Bo79}. Here $\tau_0$
is the line center optical depth from the center to the edge of the slab,
and $t_{\rm light}$ is the light crossing time if the medium were
transparent (i.e. $t_{\rm light}=R/c$ in the case of the cloud described
above). Note however, that the precise value of $t_{\rm trap}$ depends on
other factors including for example, the gas distribution (clumpiness and
geometry), the velocity distribution of the gas, and the dust content of
the cloud \citep{Bo79}.  The Ly$\alpha$ radiation
pressure becomes comparable to gravity when
\begin{equation}
L_{\alpha,{\rm 41}}=1.0\Big{(}\frac{M}{10^{8}M_{\odot}}\Big{)}^{4/3}\Big{(}\frac{16}{1+z}\Big{)}^2\Big{(}\frac{15t_{\rm light}}{t_{\rm trap}}\Big{)}, 
\label{eq:lcrit}
\end{equation} 
where $L_{\alpha}=L_{\alpha,{\rm 41}}\times 10^{41}$ erg s$^{-1}$, and
where we have substituted the virial radius of a galaxy
mass $M$, $R_{\rm vir}=0.97$
kpc $\times(M/10^8M_{\odot})^{1/3}$$(1+z/16)^{-1}$, for $R$ (Eq.~24 of Barkana \& Loeb
2001). For comparison, a star forming galaxy can generate a Ly$\alpha$
luminosity of $L_{\alpha}=(10^{42}-10^{43})\times({\rm
SFR}/M_{\odot}\hs{\rm yr}^{-1})$ erg s$^{-1}$, where the precise conversion
factor depends on the gas metallicity and the stellar initial mass function
\citep[e.g.][]{S03}. Therefore, a star formation rate of merely SFR$\gsim
0.01$--$0.1M_{\odot}$ yr$^{-1}$ is needed to generate a Ly$\alpha$
luminosity that is capable of unbinding gas from a halo of mass
$10^8-10^9M_{\odot}$.

Halos of $\la 10^9M_\odot$ are very common at $z\gsim 6$, and have a
sufficiently large reservoir of baryons to sustain the above-mentioned star
formation rates for a prolonged time. In this paper we provide a more
detailed investigation of the magnitude of Ly$\alpha$ radiation pressure in
the environment of high-redshift star forming galaxies.
In particular, we use a Ly$\alpha$ Monte-Carlo radiative transfer code
\citep{mc} to compute Ly$\alpha$ radiation pressure in a wider range of
models.
Our treatment of radiative transfer and our focus on the environment of
high-redshift star forming galaxies, distinguish this paper from previous
work. We will show that the radiation force exerted by Ly$\alpha$ photons
on neutral hydrogen gas can exceed the gravitational force that binds the gas
to its host galaxy by orders of magnitude, and may drive supersonic
outflows of neutral gas both in the intergalactic and the interstellar
medium.

The outline of this paper is as follows.
In \S~\ref{sec:code} we describe how Ly$\alpha$ radiation pressure is
computed in the Monte-Carlo radiative transfer code, and show the tests
that are performed to test the accuracy of the code.
In \S~\ref{sec:result}, we present our numerical results.  Finally,
\S~\ref{sec:conc} summarizes the implications of our work and our main
conclusions. The cosmological parameter values used throughout our
discussion are
$(\Omega_m,\Omega_{\Lambda},\Omega_b,h)=(0.27,0.73,0.042,0.70)$
\citep{Komatsu08}.

\section{Ly$\alpha$ Radiation Pressure}
\label{sec:code}

The force $F_{\rm rad}$ experienced by an atom in a direction ${\bf n}$ is
related to the flux through a plane normal to ${\bf n}$,
\begin{equation}
F_{\rm rad}=\frac{4\pi}{c}\int d\nu \hs \sigma({\nu}) H(\nu),
\label{eq:flux1}
\end{equation} 
where $\sigma(\nu)$ is the Ly$\alpha$ absorption cross-section at frequency
$\nu$. The specific flux is given by $H(\nu)=\frac{1}{2}\int d\mu\hs \mu
I(\mu,\nu)$, where $I(\nu,\mu)$ is the specific intensity of the radiation
field \citep[see, e.g. Eq. 1.113 in][]{RL79}, and $\mu={\bf n}\cdot {\bf
k}$ in which ${\bf k}$ denotes the propagation direction of the radiation
(i.e. $\mu=1$ for radiation propagating perpendicular to the plane).

The specific intensity obeys the radiative transfer equation, which reads
(in spherical coordinates)
\begin{equation}
\mu \frac{\partial I}{\partial r}+\frac{(1-\mu^2)}{r}\frac{\partial
I}{\partial \mu}=\chi_{\nu}(J-I)+S_{\nu}(r),
\label{eq:RT}
\end{equation}
where in this equation $\mu \equiv {\bf r}\cdot {\bf k}/|{\bf r}|$,
$J(\nu)=\int d\mu\hs I(\mu,\nu)$ denotes the mean intensity, and $S_{\nu}(r)$ the emission function for newly created photons at frequency $\nu$ and radius $r$ (in photons cm$^{-3}$ s$^{-1}$ sr$^{-1}$ Hz$^{-1}$, see e.g. Loeb \& Rybicki
1999). Furthermore, $\chi_{\nu}=\frac{h_P\nu_{\alpha}}{4\pi}\frac{B_{21}}{\sqrt{\pi}\Delta \nu_{\alpha}}\big{(}3n_1-n_2\big{)}\phi(\nu)$ denotes the opacity at frequency $\nu$ \citep[e.g.][]{RL79}, where $h_{\rm p}$ is Planck's constant, $\nu_\alpha=2.46\times 10^{15}$ Hz is the Ly$\alpha$ frequency, $n_{1(2)}$ is the number density of hydrogen atoms in their electronic ground (first excited) state, $B_{21}$ is the Einstein-B coefficient of the $2\rightarrow 1$ transition, $\phi(\nu)$ is the line profile function (e.g. Rybicki \& Lightman 1979, their Eq. 1.79), and $\Delta \nu_{\alpha}=\frac{v_{\rm th}}{c}$. Here, $v_{th}$ is the thermal velocity of the hydrogen atoms in the gas, given by $v_{th}=\sqrt{2k_B T/m_p}$, where $k_B$ is the Boltzmann constant, $T$ the gas temperature, and $m_p$ the proton mass.

Under the assumption that $I(\nu,\mu)$ has only a weak dependence on
direction (which is reasonable given that Ly$\alpha$ radiation scatters
very frequently), $I(\nu,\mu)$ can be expressed as a first-order Taylor
expansion in $\mu$, i.e. $I(\nu,\mu)=a(\nu)+b(\nu)\mu$. In this so-called
``Eddington approximation'', the expression for flux simplifies to
\citep[e.g.][ their Eq. 1.118]{RL79}
\begin{equation}
H(\nu)=\frac{1}{3}\frac{dJ(\nu)}{d\tau}=\frac{c}{12\pi}\frac{du(\nu)}{d\tau},
\label{eq:flux2}
\end{equation} where we have used the relation, $u=4\pi J/c$, in which $u(\nu)$ is the specific energy density in the radiation field at a frequency $\nu$. Furthermore, we have decoupled the gas' absorption and emission functions from the Ly$\alpha$ radiation field, and assumed that all neutral hydrogen atoms are in their electronic ground state (this assumption is justified in more detail in Appendix~\ref{app:n2}), i.e. $n_2=0$ and $n_1=n_H$. Under this assumption,  $d\tau=\chi_{\nu}dr=n_H\sigma(\nu)dr$ with $n_H$ being the number density of neutral hydrogen atoms. Substituting this expression back into Eq.~(\ref{eq:flux1}) yields
\begin{equation}
F_{\rm rad}=\frac{1}{3n_H}\frac{d}{dr}\int d\nu \hs
u(\nu)=\frac{1}{3n_H}\frac{dU}{dr},
\label{eq:force}
\end{equation} 
where we defined $U \equiv \int d\nu\hs u(\nu)$. Note that the
cross-section does not appear in the final expression for the radiation
force\footnote{The right-hand-side of Eq.~(\ref{eq:force}) is analogous to
the usual pressure gradient force in fluid-dynamics which is not dependent
on the scattering cross-section of the fluid particles.}.

\subsection{Implementation in Monte-Carlo Technique}
\label{sec:radpresmc}

In our Monte-Carlo simulation we sample the gas density and velocity fields
with $N_s=5000$ concentric spherical shells. The radius, thickness, and volume of
shell $j$ are denoted by $r_j$, $dr_j$, and $V_j$, respectively. We compute
the radiation force using two approaches:

\begin{itemize}

\item In the first approach, we calculate the energy density ($U$ in
Eq.~\ref{eq:force}) in the Ly$\alpha$ radiation field as a function of
radius: Using the Monte-Carlo simulation we compute the average time that
photons spend in shell $j$, which we denote by $\langle t \rangle_j$. The
total number of photons that is present in shell $j$ at any given time is
then given by $N_{\alpha,j}=\dot{N}_{\alpha}\times \langle t \rangle_j$,
where $\dot{N}_{\alpha}$ is the rate at which photons are emitted. This
yields the energy density, $U_j=N_{\alpha,j}h\nu_{\alpha}/V_j$. Finally, Eq.~(\ref{eq:force}) is used to compute the radiation force on atoms in shell $j$. Note that estimators of the energy density in -and the momentum transfer by- a radiation field in a more general context is discussed by e.g. \citet{Lucy99} and \citet{Lucy07}.

\item In the second approach, we calculate the momentum transfer from a
Ly$\alpha$ photon to an atom in each scattering event, $\Delta p=h_{\rm p}
({\bf k}_{\rm in}-{\bf k}_{\rm out})/2\pi$. Here, ${\bf k}_{\rm in}$ and ${\bf k}_{\rm out}$ are the photons
wavevectors before and after scattering. We compute the average {\it total
momentum transfer} (i.e. summed over all scattering events) per photon in
shell $j$, $\langle \Delta \mathcal{P}\rangle_j$, and obtain the total
momentum transfer from $\dot{P}_{\alpha,j}=\dot{N}_{\alpha}\times \langle
\Delta \mathcal{P}\rangle_j$. The force on an individual atom is obtained
by dividing by the total number of hydrogen atoms in shell $j$, i.e
$F_j=\dot{P}_{\alpha,j}/(V_j\times n_{H,j})$.

\end{itemize}

Both methods should give identical results, provided that the Eddington
approximation holds.

\subsection{Test Case: Sources in a Neutral Comoving IGM}
\label{sec:test}

We begin by considering a Ly$\alpha$ point source at a redshift $z=10$
embedded in a neutral intergalactic medium (IGM) that is expanding with the
Hubble flow. The photons scatter and diffuse away from the source while
Hubble expansion redshifts the photons away from resonance. In this case, the angle-averaged intensity $J(\nu)$ and its radial dependence can be
calculated analytically \citep{LR99}. The availability of analytic
expressions for $J(\nu,r)$, and therefore the radiation force $F_{\rm rad}$
(through Eq.~\ref{eq:force}), makes this a good test case for our code.

\begin{figure}
\vbox{\centerline{\epsfig{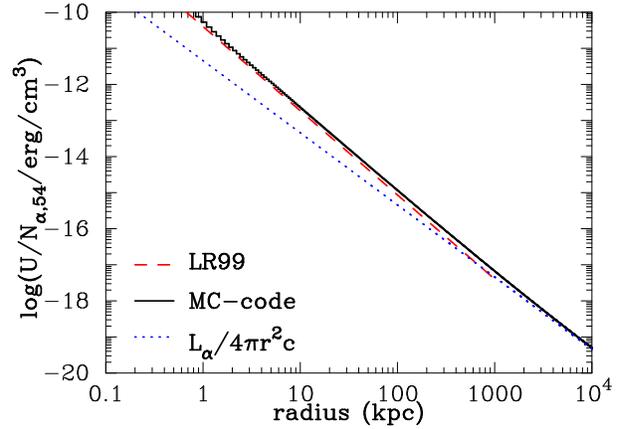}}}
\caption[]{Radial profile of the energy density $U(r)$ (erg cm$^{-3}$) in
the Ly$\alpha$ radiation field surrounding a central source that is
emitting $\dot{N}_{\alpha,54} \times 10^{54}$ photons s$^{-1}$ into an expanding neutral IGM. The {\it blue dotted line} shows $U(r)$ if the IGM were fully transparent. The {\it black solid
(red dashed) line} shows $U(r)$ when radiative transfer is included using an
analytic (Monte-Carlo) approach (see text). Scattering reduces the speed at
which Ly$\alpha$ photons are propagating radially outward, increasing
$U(r)$ relative to the transparent case.}
\label{fig:uden}
\end{figure}

In Figure~\ref{fig:uden} we plot the radial dependence of the energy
density (in erg cm$^{-3}$) in the Ly$\alpha$ radiation field for a
model in which the central source is emitting
$\dot{N}_{\alpha,54}\times 10^{54}$ photons s$^{-1}$ (where we have
introduced the dimensionless quantity $\dot{N}_{\alpha,54}\equiv
(\dot{N}_{\alpha}/10^{54}$ photons s$^{-1}$). This corresponds to a
luminosity of $L_{\alpha}=\dot{N}_{\alpha,54} \times 1.6\times
10^{43}$ erg s$^{-1}$, which represents a bright Ly$\alpha$ emitting
galaxy \citep[e.g.][]{Ouchi08}. The {\it blue dotted line} shows the
energy density if the IGM were fully transparent to Ly$\alpha$
radiation. In this hypothetical case all photons stream radially
outward, and the energy density is given by $L_{\alpha}/(4\pi r^2
c)$. The {\it red dashed line} shows the energy density,
$U(r)=\frac{4\pi}{c}\int d\nu\hs J(\nu,r)$, derived from the analytic
expression for $J(r,\nu)$ given in Loeb \& Rybicki (1999, their
Eq.~21), while the {\it black histogram} shows the energy density
extracted from the simulation (\S~\ref{sec:radpresmc}). Clearly, the
analytic and Monte-Carlo calculations yield consistent
results. Scattering reduces the effective speed at which photons
propagate radially outward, which enhances their energy density
(especially at small radii) relative to the transparent case. At
sufficiently large distances however, the photons have redshifted far
enough from resonance that they are propagating almost freely to the
observer, and the energy density approaches $L_{\alpha}/4\pi r^2
c$. We note that at a sufficiently high value of
$\dot{N}_{\alpha,54}$, the fraction of hydrogen atoms that populate
the 2p (and 2s) levels is non-negligible and our assumption that (almost) all
of the atoms populate their electronic ground state becomes invalid
(so that the solution for $U(r)$ in Figure~\ref{fig:uden} breaks
down). However, as we show in Appendix~\ref{app:n2}, this only occurs
when $\dot{N}_{\alpha,54} \gsim 10^7$, well beyond the regime
considered in this paper.

\begin{figure}
\vbox{\centerline{\epsfig{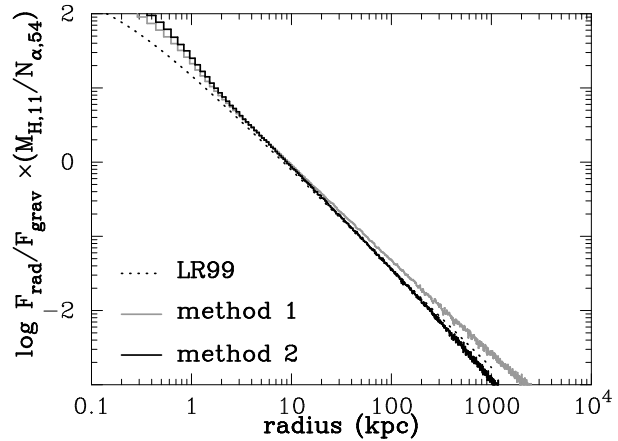}}}
\caption[]{The ratio of radiation to gravitational force on a hydrogen atom
as a function of physical radius in kpc. To scale out the dependence of
this ratio on halo mass, $M_h$, and production rate of Ly$\alpha$ photons
by the central source, $\dot{N}_{\alpha}$, the vertical axis is normalized
by $M_{h,11}/\dot{N}_{\alpha,54}$ (see text). The {\it black dotted (grey solid) line}
was obtained by applying Eq.~(\ref{eq:force}) to the energy density $U(r)$
that was obtained by using the analytic (Monte-Carlo) approach (see
Fig~\ref{fig:uden}). The {\it black solid line} was obtained by directly computing
the momentum transfer rate from photons to atoms in the Monte-Carlo code as
outlined in \S~\ref{sec:radpresmc}. The results demonstrate that ({\it i})
the radiation force exceeds gravity at $r<10
(\dot{N}_{\alpha,54}/M_{h,11})$ kpc, and ({\it ii}) both methods yield
consistent results.}
\label{fig:force1}
\end{figure}
In Figure~\ref{fig:force1} we compare the radiation force to the
gravitational force on a single hydrogen atom, $F_{\rm grav}=GM(<r)m_p/r^2$,
where $M(<r)$ is the total mass enclosed within a radius $r$). We plot the
ratio $F_{\rm rad}/F_{\rm grav}$ scaled by $M=10^{11}M_{\odot}$
\footnote{The number density of halos more massive than $10^{11}M_{\odot}$
at $z=10$ is $\sim 10^{-7}$ comoving Mpc$^{-3}$, implying that these rare
halos are among the most massive ones in existence at that early cosmic
time.}. The {\it black dotted line (grey solid histogram)} was calculated by applying
Eq.~(\ref{eq:force}) to the energy density $U(r)$ that was obtained by
using the analytic (Monte-Carlo) approach (also see
Fig~\ref{fig:uden}). For comparison, the {\it black solid histogram} was obtained by
directly computing the momentum transfer rate from photons to atoms as
outlined in \S~\ref{sec:radpresmc}. Figure~\ref{fig:force1} shows that the
radiation force overwhelms gravity at small radii. The energy density
scales approximately as $\partial \log U/\partial \log r \sim-2.3$
(Fig~\ref{fig:uden}). Therefore, $F_{\rm rad}/F_{\rm grav} \propto r^{-1.3}$ and
reaches unity at $r \sim 10$ physical kpc.

The radiation force increases linearly with $\dot{N}_{\alpha}$ while the
gravitational force scales as $M$. Thus, $F_{\rm rad}/F_{\rm grav}$ scales linearly
with the ratio $\mathcal{R}\equiv \dot{N}_{\alpha}/M$. To scale out the
dependence on $\mathcal{R}$, the vertical axis shows the quantity
$({F_{\rm rad}}/{F_{\rm grav}})\times ({M_{11}}/{\dot{N}_{\alpha,54}})$, where $M_{11}=(M/10^{11} M_{\odot})$. For example, if $M_{11}=0.1$ then
Figure~\ref{fig:force1} shows that radiation pressure exceeds gravity
out to $r=100$ kpc, well beyond the virial radius of a halo of this mass at $r_{\rm vir}\sim 6.6$ kpc.

Most importantly, Figure~\ref{fig:force1} shows that the two approaches
used to compute the radiation force in the simulation yield consistent
results, with a noticeable deviation only at the largest radii ($r\sim 1$
Mpc). At large radii most photons stream outwards radially and the
Eddington approximation that was used to derive Eq.~(\ref{eq:flux2})
becomes increasingly unreliable.

Next, we use the radiative transfer code to explore the magnitude of the
Ly$\alpha$ radiation pressure for a range of models which represent an
evolutionary sequence of structure formation in the Universe. We focus on
the Ly$\alpha$ radiation pressure on gas surrounding ({\it i}) the first
stars (\S~\ref{sec:firststar}); ({\it ii}) the first galaxies
(\S~\ref{sec:firstgalaxy}); and ({\it iii}) the interstellar medium of
galaxies (\S~\ref{sec:wind}).

\section{Results}
\label{sec:result}

\subsection{Case I: A Single Massive Star in a Minihalo}
\label{sec:firststar}
\begin{figure*}
\vbox{\centerline{\epsfig{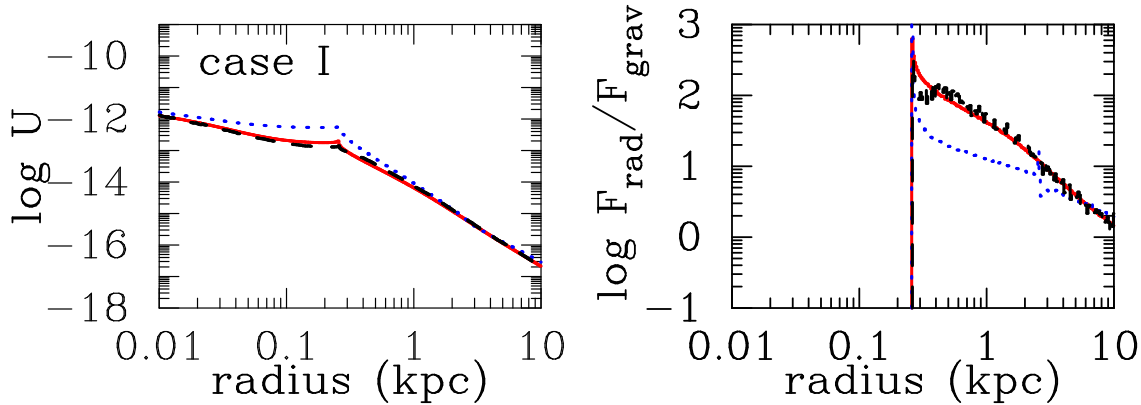}}}
\caption[]{The energy density in the Ly$\alpha$ radiation field ({\it
left panel}), and the ratio between the radiation and the
gravitational forces ({\it right panel}) for a single very massive
star ($M_*=100M_{\odot}$) in a minihalo $M_h=2\times 10^6M_{\odot}$
. The {\it red solid} line represents a model in which the IGM is
at the mean cosmic density and undergoes Hubble expansion right
outside the virial radius at $r=0.26$ kpc. The {\it blue dotted line}
shows a more realistic model in which the IGM is overdense near the
virial radius, and in which the intergalactic gas is gravitationally
pulled towards the minihalo (see text). The {\it black dashed line}
shows the same model as the {\it red solid line} but with the neutral
fraction increasing linearly between $r_{\rm vir}$ and $2r_{\rm
vir}$. The star ionizes all the gas out to the virial
radius. Ly$\alpha$ photons freely propagate until they reach the edge
of the HII region, where they are likely to be scattered back into the
ionized minihalo. This yields an almost constant radiation energy
density. Once outside the HII region, the radiation force dominates
over gravity out to $r=10$ kpc and may accelerate neutral gas outside
the HII region to velocities of order $\sim 10$ km s$^{-1}$.}
\label{fig:caseI}
\end{figure*}
Numerical simulations of structure formation suggest the first stars that
formed in our Universe were massive ($M_\star \sim 100M_{\odot}$), and
formed as single objects in dark matter halos with masses of $M\sim
10^6M_{\odot}$ that collapsed at $z>10$
\citep[e.g.][]{Haiman96,Abel02,Yoshida06}. Here, we focus our attention on
a star with a mass $M_*=100M_{\odot}$ that formed at $z=15$ in a dark
matter mini-halo of mass $M=2\times 10^6M_{\odot}$. The star emits
$10^{50}$ ionizing photons per second \citep{S02,Abel07}. We assume that
the ionizing flux ionizes all the gas out to the virial radius of the dark
matter halo ($r_{\rm vir}=0.26$ kpc) but not beyond that radius
\citep{Kita04}. Hence, the IGM gas surrounding this central source is
assumed to be neutral ($x_{\rm HI}=1.0$) and cold ($T_{\rm gas}=300$K, which
corresponds to the temperature of the neutral IGM at $z=15$ due to X-Ray heating, see e.g. Fig~1 of Pritchard \& Loeb 2008).
 
Recombination following photoionization converts $\sim 68\%$ of all
ionizing photons into Ly$\alpha$ photons \citep[][p
387]{Osterbrock89}. Hence, the entire halo is a Ly$\alpha$ source that is
surrounded by neutral intergalactic gas. To determine the radial dependence
of the Ly$\alpha$ production rate ($S_{\nu}$ in Eq.~\ref{eq:RT}), we need
to specify the gas density profile. We assume that the gas distribution
inside the dark matter halo is described by an NFW-profile with a
concentration parameter $C=5$ and a thermal core\footnote{With this
gas density profile, the total recombination rate inside the dark matter
halo is $\int_0^{r_{\rm vir}}dr\hs 4\pi r^2n_{\rm H}^2\alpha_{\rm rec}\sim
4.5\times 10^{49}$ s$^{-1}$. The total recombination rate can be increased
to balance the photoionization rate by introducing a clumping factor
$K\equiv \langle n_H^2 \rangle/\langle n_{\rm H} \rangle^2\sim 2$.} at
$r<3r_{\rm vir}/4C$ \citep[see][]{Maller04}. We point out however, that our
final results are not sensitive to our choice of $S_{\nu}(r)$.

Once $S_{\nu}(r)$ has been determined, we find the radius, $r$, at which a
Ly$\alpha$ photon is generated in the Monte-Carlo simulation from the
relation
\begin{equation}
R=\frac{1}{N}\int_0^{r}dr\hs 4\pi r^2 n_{\rm H}^2\alpha_{\rm rec},
\label{eq:s}
\end{equation} 
where $R$ is a random number between $0$ and $1$, $N=\int_0^{r_{\rm
vir}}dr\hs 4\pi r^2 n_{\rm H}^2\alpha_{\rm rec}$ is the total recombination
rate inside the dark matter halo, and $\alpha_{\rm rec}=2.6\times 10^{-13}$
cm$^3$ s$^{-1}$ is the case-B recombination coefficient at a temperature
$T=10^4$ K \citep[e.g.][]{Hui97}. Once the photon is generated, it scatters
through the neutral IGM until it has redshifted far enough from resonance
that it can escape to the observer.

In the {\it left panel} of Figure~\ref{fig:caseI} we show the energy
density (in erg cm$^{-3}$) of the Ly$\alpha$ radiation field as a function
of radius. The {\it red solid} line represents a model in which we assumed
the IGM to follow the mean density and Hubble expansion right outside the
virial radius. The {\it blue dotted line} shows a more realistic model in
which the IGM is still overdense near the virial radius, and in which the
intergalactic gas is gravitationally pulled towards the minihalo (see
Dijkstra et al 2007 for a quantitative description of the density and
velocity profiles based on the model of Barkana 2004). The {\it black
dashed line} shows the same model as the {\it red solid line} but with the neutral fraction increasing
linearly between $r_{\rm vir}$ and $2r_{\rm vir}$. This provides a better
representation of the fact that the central population III star emits
ionizing photons with energies $\gsim 54$ eV, which can photoionize
hydrogen (and helium) atoms that lie deeper in the IGM. The goal of this
model is to investigate whether our results depend sensitively on the
presence of a sharp boundary between HI and HII.

All models show that the radiation energy density within the fully ionized
minihalo ($r\lsim r_{\rm vir}=0.26$ kpc) has only a weak dependence on
radius, i.e. $d\log U/d\log r \gsim -1$. Naively, this may appear
surprising given the fact that within the model, no scattering occurs
within the virial radius and one may expect the energy density in the
Ly$\alpha$ radiation field to scale as $U \propto r^{-2}$. However, in
reality $U$ obtains only a weak radial dependence because the radiation can be
scattered back into the ionized minihalo as soon as it 'hits' the wall of
neutral IGM gas. Ly$\alpha$ photons are therefore trapped inside the
ionized minihalo and their energy density is boosted to a value that is
only weakly dependent on radius. On the other hand, for $r\gsim r_{\rm
vir}$ we find that $d\log U/d\log r \lsim -2$, which is because Ly$\alpha$
photons are trapped more efficiently near the edge of the HII region, while
they stream freely outwards at larger radii (as in \S~\ref{sec:test} and
Fig~\ref{fig:uden}). Figure~\ref{fig:caseI} shows clearly that the radial
dependence of the Ly$\alpha$ energy density is not sensitive to the
detailed model assumptions about the gas in the IGM.

In the {\it right panel} of Figure~\ref{fig:caseI} we show the ratio
between the radiation force (Eq.~\ref{eq:force}) and the gravitational
force on a single hydrogen atom), $F_{\rm grav}={GM(<r)m_{\rm
p}}/{r^{2}}$, where $M(<r)$ is the total (baryons + dark matter) mass
enclosed within a radius $r$. In all models, radiation pressure dominates
over gravity by as much as $\gsim 2$ orders of magnitude. The radiation
force is largest for the models in which the IGM is assumed to be at
mean density, because of the $n_{\rm H}^{-1}$ factor in the
equation for the radiation force (Eq.~\ref{eq:force}). Note that the spike
near $r\sim 2.6$ kpc for the other two models is due to an artificial
discontinuity in the IGM velocity field that exists in this model. 

\begin{figure*}
\vbox{\centerline{\epsfig{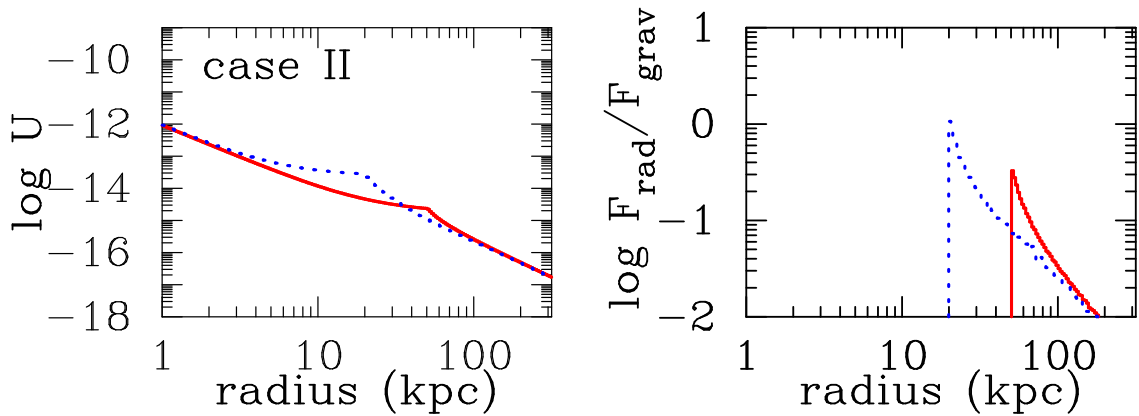}}}
\caption[]{Same as Figure~\ref{fig:caseI}, but for the case of a star
forming galaxy ($\dot{M}_*=0.34M_{\odot}$ yr$^{-1}$, see text), surrounded
by an HII region with a radius $R_{\rm HII}=50$ kpc ($R_{\rm HII}=20$ kpc)
for the {\it solid line} ({\it dotted line}), which is in turn surrounded
by a fully neutral intergalactic medium (IGM). For the assumed total halo
mass of $M_{\rm tot}=10^9M_{\odot}$, the pressure exerted by the Ly$\alpha$
photons is not large enough to exceed gravity. However, radiation pressure
wins for $M_{\rm tot}=10^8M_{\odot}$, but even in this case the radiation
force is not large enough to produce a significant wind speed in the IGM
(see text). }
\label{fig:caseII}
\end{figure*}

Ly$\alpha$ radiation pressure may operate throughout the lifetime of the
central star. Over a lifetime of $\sim 2.5$ Myr \citep[see Table 4
of][]{S02}, this mechanism is capable of accelerating the gas to velocities
of $10$ ($50$) km s$^{-1}$ at $r=0.3$ kpc in the model represented by the
{\it blue dotted} ({\it red solid}) {\it line}, and to $4$ ($16$) km
s$^{-1}$ at $r=0.4$ kpc (the reason for this large difference is that the edge of the HII region lies at $r=0.26$ kpc. Hence, gas at $r=0.3$ kpc is separated by 0.04 kpc from this edge, while gas at $r=0.4$ kpc is separated by a distance that is 3.5 times larger). 
 
Thus, Ly$\alpha$ radiation pressure can accelerate the gas to velocities
that exceed the escape velocity from the dark matter halo ($v_{\rm esc}\sim
\sqrt{2}v_{\rm circ}\sim 8$ km s$^{-1}$) as well as the sound speed of the
intergalactic medium ($c_{\rm s}=2.2(T_{\rm gas}/300\hs{\rm K})^{1/2}$ km
s$^{-1}$). 

Note the as the gas is pushed out and its velocity profile changes, the
subsequent radiative transfer is altered. For example, we repeated the
radiative transfer calculation for models in which gas at $r_{\rm vir}< r
\lsim 2r_{\rm vir}$ was accelerated to velocities in the range $10-20$ km
s$^{-1}$ (outward) and found a slightly shallower profile for $U(r)$ which lowered
the radiation force by a factor of $\sim 3$. Consequently, the acceleration
of the gas decreases with time and the actual velocities reached by the gas
are lower than the estimates given above by a factor of a few.
Nevertheless, the resulting velocities are still substantial.

Our calculations imply that Ly$\alpha$ radiation pressure can affect the
gas dynamics in the IGM surrounding minihalos that contain the first
stars. The impact of Ly$\alpha$ radiation pressure increases with
decreasing density of the surrounding gas in the IGM. In practice, the
distribution of the IGM is not spherically symmetric. Instead, the density
is expected to vary from sightline to sightline (being large along
filaments and small along voids). Our results imply that Ly$\alpha$
radiation pressure will be most efficient in 'blowing out' the lower
density gas. This conjecture is supported by the tendency of Ly$\alpha$
photons to preferentially scatter through the low-density gas; their
propagation along the path of least resistance would naturally boost up the
Ly$\alpha$ flux there. This effect will be moderated by the tendency of the
HII region around the first stars to extend further into the low density
gas (in 'butterfly'-like patterns, e.g. Abel et al, 1999).

If the central star dies in a supernova explosion, then the resulting
violent outflow could blow most of the baryons out from the
minihalo. However, stars with masses in the range $30M_{\odot}\lsim
M_{*}\lsim140M_{\odot}$ and $M_*\gsim 260M_{\odot}$, are not expected to
end their lives in a supernova. Instead, these stars collapse directly to a
black hole \citep{HW02} and have weak winds (because of the lack of
heavy elements in their atmosphere), so that radiation pressure may
be the dominant process that affects their surrounding IGM.

In summary, Ly$\alpha$ radiation pressure on the neutral IGM around
minihalos in which the first stars form, can exceed gravity by orders of
magnitude and launch supersonic winds. Our limited analysis does not allow
a detailed discussion on the consequences of these winds. This requires 3D
simulations with cosmological initial conditions that capture the full IGM
density field around the minihalo and that track the evolution of the
shocks that may form in the IGM. Such simulation are numerically
challenging as they require self-consistent treatment of gas dynamics and
Ly$\alpha$ radiative transfer in a moving inhomogeneous medium.

\subsection{Case II: A Young Star Forming Galaxy}
\label{sec:firstgalaxy}

\begin{figure*}
\vbox{\centerline{\epsfig{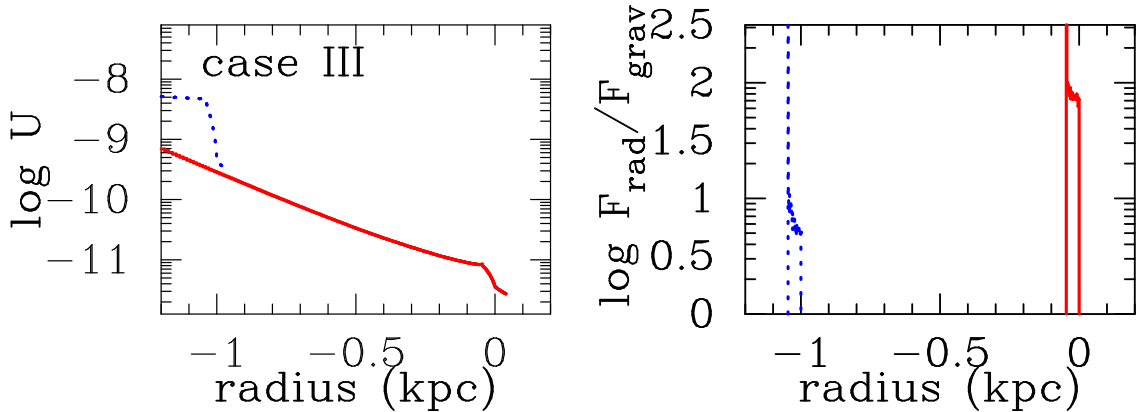}}}
\caption[]{Same as in Figure~\ref{fig:caseI} and Figure~\ref{fig:caseII},
but for models in which a central Ly$\alpha$ source of luminosity
$L_{\alpha}$ is surrounded by a thin ($r_{\rm sh,min}=0.9r_{\rm sh,max}$)
spherical shell of HI that is expanding at $v_{\rm sh}=200$ km
s$^{-1}$. The {\it blue dotted lines} ({\it red solid lines}) represent a
model in which $N_{\rm HI}=10^{21}$ cm$^{-2}$ and $r_{\rm sh,max}=0.1$ kpc
($N_{\rm HI}=10^{19}$ cm$^{-2}$ and $r_{\rm sh,max}=1.0$ kpc). When
calculating $F_{\rm grav}$, we assumed the total mass
enclosed by the supershell to be $10^8M_{\odot}$ (see text). The results
strongly suggest that Ly$\alpha$ radiation pressure may be dynamically
important in the interstellar medium of galaxies.  The {\it total}
radiation force on the shell (obtained as a sum over all atoms) may be
computed via $F_{\rm tot}=M_F L_{\alpha}$/c, where $M_F$ may be thought of
as a force multiplication factor that depends both on the shell's outflow
speed, $v_{\rm exp}$, and its HI column density, $N_{\rm HI}$. The
dependence of $M_F$ on these parameters is shown in Fig~\ref{fig:fscat}.}
\label{fig:caseIII}
\end{figure*}

Our second case concerns a young galaxy that is forming multiple stars in a
dark matter halo of mass $M=10^9 M_{\odot}$ at $z=10$. We assume that the
galaxy is converting a fraction $f_*=10\%$ of its baryons into stars over
$\sim 0.1t_{\rm H}$ \citep[][]{Wyithe06}, where $t_{\rm
H}=[{2}/{3H(z)}]\sim0.49$ Gyr, is the age of the Universe at $z=10$. This
translates to a star formation rate of $\dot{M}_*=0.34M_{\odot}$
yr$^{-1}$. For population III stars forming out of pristine gas, the total
emission rate of ionizing photons is $\dot{N}_{\rm ion}\sim 3\times
10^{53}$ s$^{-1}$ \citep{S02}\footnote{More precisely, the ionizing photon
production rate is $\sim 10^{54}$ s$^{-1}$ $\times({\rm
SFR}/M_{\odot}\hs{\rm yr}^{-1})$ in the no-mass-loss model of Schaerer
(2002) in which metal-free stars form according to a Salpeter IMF with
$M_{\rm low}=1M_{\odot}$ and $M_{\rm high}=500 M_{\odot}$ (his model
'B').}. If $\sim 1\%$ of the ionizing photons escape from the galaxy
\citep[][]{Chen07,Gnedin08}, then this translates to a Ly$\alpha$
luminosity of $L_{\alpha}=3\times 10^{42}$ erg s$^{-1}$. Furthermore, this
galaxy can photoionize a spherical HII region of a radius $R_{\rm HII}\sim
50$ physical kpc. Note however, that other ionizing sources
would likely exist within this HII region. Indeed, clusters of sources are
thought to determine the growth of ionized bubbles during
reionization. This results in a characteristic HII region size that is
significantly larger than that produced by single source, especially during
the later stages of reionization \citep[e.g.][]{F04,McQuinn07}. In this
framework, our model represents a star forming galaxy during the early
stages of reionization or alternatively a galaxy that lies $50$ kpc away
from the edge of a larger ionized bubble.

In this particular case, the majority of all recombination events occur in
the central galaxy. Thus, we initiate all Ly$\alpha$ photons at $r=0$ in
the Monte-Carlo simulation. We assume that the gas is completely ionized
out to $R_{\rm HII}=50$ kpc, beyond which it is neutral. As 
shown in \S~\ref{sec:firststar}, this abrupt transition in the ionized
fraction of H in the gas does not affect our results.

The {\it left panel} of Figure~\ref{fig:caseII} shows the energy density
(in erg cm$^{-3}$) of the Ly$\alpha$ radiation field as a function of
radius. The {\it solid line} represents the model discussed above. A kink
in the energy density is seen at the edge of the HII region (see
\S~\ref{sec:firststar} for a more detailed discussion of the profile). The
{\it dotted line} represents a variant of the model in which we have
reduced the size of the HII region to $R_{\rm HII}=20$ kpc.
 
The {\it right panel} of Figure~\ref{fig:caseII} shows the ratio between
the radiation and the gravitational forces on a single hydrogen atom. In
our fiducial model, the radiation force does not exceed gravity; rather, at
the edge of the HII region, gravity is $\sim 3$ times stronger. The
radiation force becomes equal to the gravitational force if $R_{\rm
HII}=20$ kpc. This requires an extremely low [by a factor $\sim (50/20)^3$]
escape fraction of ionizing photons, $f_{\rm esc}\sim 6\times10^{-4}$.

Alternatively, radiation pressure {\it is} important when the halo mass of
the star forming region is reduced to $10^8M_{\odot}$. Halos of this mass are the the most abundant halos at $z\sim 10$ that are capable of cooling via excitation of atomic hydrogen (i.e. their virial
temperature just exceeds $T_{\rm vir}\sim 10^4$ K, e.g. Barkana \& Loeb
2001). The total gas reservoir inside these halos is $M_{\rm
b}=\frac{\Omega_{\rm b}}{\Omega_{\rm m}}M_{\rm tot}\sim 1.5\times 10^7
M_{\odot}$, and so these halos can sustain a star formation rate of
$\dot{M}_{*}=0.3M_{\odot}$ yr$^{-1}$ for up to $\sim 50$ Myr. However, even
if the radiation force is allowed to operate for $\sim 50$ Myr, we find
that radiation pressure cannot accelerate the gas in the IGM to velocities
that exceed $\sim 1$ km s$^{-1}$. We therefore conclude that although
Ly$\alpha$ pressure may exceed gravity in the neutral IGM that surrounds
HII regions around $M_{\rm tot}=10^8M_{\odot}$ halos, the absolute
magnitude of the radiation force is too weak to drive the IGM to supersonic
velocities.

\subsection{Case III: Ly$\alpha$ Driven Galactic Supershells}
\label{sec:wind}

In principle, Ly$\alpha$ radiation pressure can be important when neutral
gas exists in close proximity to a luminous Ly$\alpha$ source. So far, we
focused our attention on HI gas in the IGM. However, neutral gas in the
interstellar medium (ISM) of the host galaxy is located closer to the
Ly$\alpha$ sources and should be exposed to an even stronger Ly$\alpha$
radiation pressure.  Indeed, it has been demonstrated
\citep[e.g.][]{Ahn02,Verhamme08} that scattering of Ly$\alpha$ photons by
neutral hydrogen atoms in a thin (with a thickness much smaller than its radius), outflowing 'supershell' of HI gas surrounding
the star forming regions can naturally explain two observed phenomena:
({\it i}) the common shift of the Ly$\alpha$ emission line towards the red
relative to metal absorption lines and the host galaxy's systemic redshift determined from other nebular recombination lines \citep[e.g.][]{Pettini01,Shapley03}; and ({\it ii}) the asymmetry of the Ly$\alpha$ line with emission extending well into its red wing \citep[e.g.][]{L95,Tapken07}.

The existence of thin, outflowing shells of neutral atomic hydrogen around HII regions is confirmed by
HI-observations of our own Milky-Way \citep[][]{Heiles84} and other nearby
galaxies \citep[e.g.][]{Ryder95}. The largest of these shells, so-called
'supershells', have radii of $r_{\rm max}\sim 1$ kpc
\citep[e.g][]{Ryder95,M02} and HI column densities in the range $N_{\rm
HI}\sim 10^{19}$--$10^{21}$ cm$^{-2}$
\citep[e.g.][]{L95,Kunth98,Verhamme08}. Supershells are thought to be
generated by stellar winds or supernovae explosions which sweep-up gas into
a thin expanding neutral shell \citep[see e.g.][for a review]{T88}. The
back-scattering mechanism attributes both the redshift and asymmetry of the
Ly$\alpha$ line to the Doppler boost that Ly$\alpha$ photons undergo as
they scatter off the outflow on the far side of the galaxy back towards the
observer \citep[e.g.][]{Lee98,Ahn02,Ahn03,Ahn04,Verhamme06,Verhamme08}. It is interesting to investigate whether Ly$\alpha$ radiation
pressure may provide an alternative mechanism that determines the
supershell kinematics.

In Figure~\ref{fig:caseIII} we show the energy density ({\it left panel})
and the Ly$\alpha$ radiation force ({\it right panel}) for two models. Both
models assume that: ({\it i}) there is a Ly$\alpha$ source at $r=0$ with a
luminosity of $L_{\alpha}=10^{43}$ erg s$^{-1}$; ({\it ii}) the emitted
Ly$\alpha$ spectrum prior to scattering has a Gaussian shape as a function
of photon frequency with a Doppler velocity width of $\sigma=50$ km
s$^{-1}$; ({\it iii}) the spatial width of the supershell is $10\%$ of its
radius; and ({\it iv}) the shell has an outflow velocity of $v_{\rm}=200$
km s$^{-1}$. The {\it blue dotted (red solid) lines} represent a model in
which the supershell has a column density of $N_{\rm HI}=10^{21}$
($N_{\rm HI}=10^{19}$) cm$^{-2}$ and a maximum radius that is $r_{\rm
sh}=0.1$ kpc ($r_{\rm sh}=1.0$) kpc. Our calculations assume that there is no neutral gas (or dust) interior to the HI supershell.
\begin{figure}
\vbox{\centerline{\epsfig{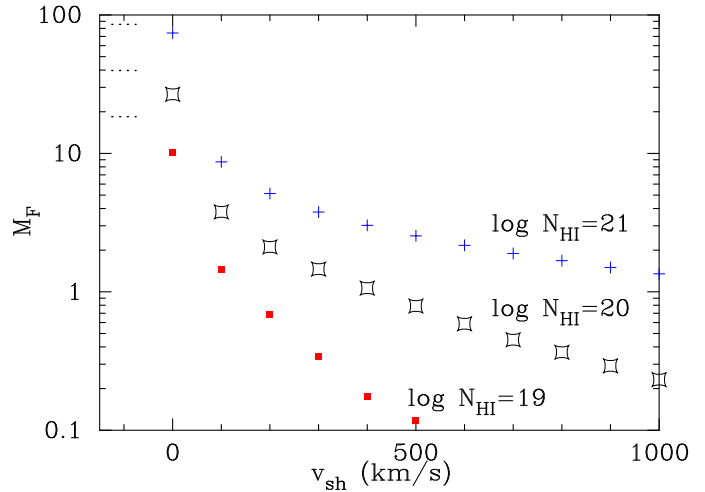}}}
\caption[]{The multiplication factor $M_F$ provides the total force that
Ly$\alpha$ photons exert on a spherical HI shell, $F_{\rm tot}=M_F
L_{\alpha}/c$, where $L_{\alpha}$ is the Ly$\alpha$ luminosity of the
source in erg s$^{-1}$ and $c$ is the speed of light. The plot shows $M_F$
as a function of the expansion velocity of the HI shell, $v_{\rm sh}$, for
three values of HI column density, and under the assumption that the HI shell surrounds an empty cavity. The parameter $M_F$ provides a measure
of the efficiency by which Ly$\alpha$ photons can be 'trapped' by the shell
of HI gas. Thus, $M_F$ increases with increasing $N_{\rm HI}$ and
decreasing $v_{\rm sh}$. The {\it dotted horizontal lines} show the values of
$M_F$ that have been derived in the past (for a static, uniform,
infinite slab of material e.g. Adams 1975).}
\label{fig:fscat}
\end{figure}

The {\it left panel} of Figure~\ref{fig:caseIII} shows that inside the
supershell the energy density decreases more gradually than $r^{-2}$
because of photon trapping (similarly to the previously discussed cases in
\S~\ref{sec:firststar}-\S~\ref{sec:firstgalaxy}). The shell with the larger
column of HI is more efficient at trapping the Ly$\alpha$ photons, and thus
yields a flatter energy density profile. In both models the energy density
drops steeply within the supershell (the energy density decreases as $r^{-2}$ outside the shell, if no scattering occurs here).

The {\it right panel} of Figure~\ref{fig:caseIII} shows the ratio between
the radiation and gravitational forces.  Towards the center of the dark
matter halo, baryons dominate the mass density and an evaluation of $F_{\rm
grav}$ requires assumptions about the radial distribution of the
baryons. For simplicity, we consider a fixed total mass interior to the
supershell of $M(<r)=10^8M_{\odot}$, so that $F_{\rm
grav}(r)=GM(<r)m_{p}/r^{2}$. Note that any assumed mass profile $M(<r)$
will not affect the results as long as $F_{\rm rad}\gg F_{\rm grav}$.

We find that the radiation force exceeds gravity in both examples under
consideration. For $N_{\rm HI}=10^{21}$ cm$^{-2}$ and $r_{\rm sh}=0.1$ kpc,
$F_{\rm rad}\sim 10 F_{\rm grav}$. Thus, radiation pressure would have been
important even if we had chosen $M\sim 10^{9}M_{\odot}$. 
In the model with $N_{\rm HI}=10^{19}$ cm$^{-2}$ and $r_{\rm sh}=1.0$ kpc,
$F_{\rm rad}\sim 10^2 F_{\rm grav}$, and radiation pressure would have been
important even if $M\sim 10^{10}M_{\odot}$. Hence, our calculations
strongly suggest that Ly$\alpha$ radiation pressure may be dynamically
important in the ISM of galaxies.

The {\it total} Ly$\alpha$ radiation force is obtained by summing the force
over all atoms in the supershell. It is interesting to compare this force
to $L_{\alpha}/c$. The latter quantity denotes the total momentum transfer
rate (force) from the Ly$\alpha$ radiation field to the supershell under
the assumption that each Ly$\alpha$ photon is re-emitted isotropically
after entering the shell (including multiple scatterings inside the
supershell). In Figure~\ref{fig:fscat} we plot the quantity $M_F$ which is
defined as
\begin{equation}
M_F\equiv \frac{\sum_{\rm atoms}F_{\rm rad}}{L_{\alpha}/c},
\end{equation} 
as a function of the expansion velocity of the shell, $v_{\rm sh}$ for
three different values of $N_{\rm HI}$.

\begin{figure}
\vbox{\centerline{\epsfig{file=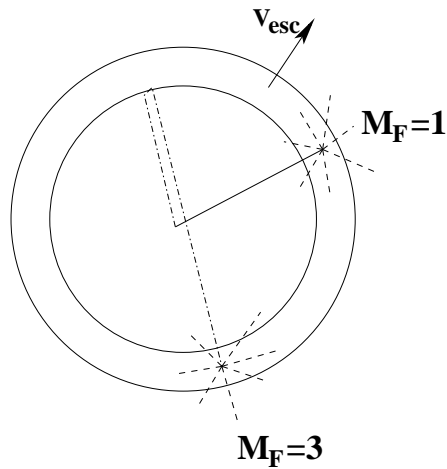,angle=0,width=6.0truecm}}}
\caption[]{A schematic illustration of the origin of the force
multiplication factor $M_F$ in an expanding supershell. The Ly$\alpha$
source is located at the center of the expanding HI supershell. The {\it
solid line} represents the trajectories of photons that enter the shell,
scatter once, and then escape from the shell isotropically (indicated by
the {\it dashed lines}). This corresponds to $M_F=1$. On the other hand,
the {\it dot-dashed line} represents the trajectories of photons that enter
the shell, are scattered back in the opposite direction, and escape with no
preferred direction after scattering in the shell for a second time. This
corresponds to a case with $M_F=3$. In general, $M_F=t_{\rm trap}/(r_{\rm
sh}/c)$.}
\label{fig:scheme}
\end{figure}

Figure~\ref{fig:fscat} shows that $M_F$, which can be thought of as a
force multiplication factor\footnote{This term derives from the
(time-dependent) force-multiplication function $M(t)$ that was
introduced by \citet{Castor75}, as $F_{\rm rad}\equiv
M(t)({\tau_eL_{\rm bol}}/{c})$. Here, $F_{\rm rad}$ is the total
force that radiation exerts on a medium, $\tau_e$ is the total optical
depth to electron scattering through this medium. The function $M(t)$
arises because of the contribution of numerous metal absorption lines
to the medium's opacity, and can be as large as $M_{\rm max}(t)\sim
10^3$ in the atmospheres of O-stars \citep{Castor75}.}, greatly
exceeds unity for low shell velocities and large HI column
densities. The parameter $M_F$ is related to the mean number of times
that a Ly$\alpha$ photon 'bounces' back and forth between opposite
sides of the expanding shell. For example, $M_F=1$ when all Ly$\alpha$
photons enter the shell, scatter once, and then escape from the shell
in no preferred direction. On the other hand, $M_F=3$ when all
Ly$\alpha$ photons enter the shell, scatter back towards the opposite
direction, and then escape in no preferred direction after scattering
in the shell for a second time. A schematic illustration of this
argument is provided in Figure~\ref{fig:scheme}. Note that when
$M_F=1$ ($M_F=3$), each photon spends on average a timescale of
$r_{\rm sh}/c$ ($3r_{\rm sh}/c$) in the bubble enclosed by the
shell. In other words, the factor $M_F$ relates to the 'trapping
time', $t_{\rm trap}$, that denotes the total time over which
Ly$\alpha$ photons are trapped inside the supershell\footnote{This
argument ignores the time spent on scattering inside the supershell
itself. Photons penetrate on average an optical depth $\tau=1$ into
the shell. If this corresponds to a physical distance that is
significantly smaller than the thickness of the shell (denoted by
$\Delta r_{\rm sh}$), then only a tiny fraction of the photons will
diffuse through the shell. Hence, when averaged over these photons,
ignoring the time spent inside the supershell itself is
justified. Alternatively, photons with a mean free path that is at
least comparable to the thickness of the shell, only spend a time
$\sim \Delta r_{\rm sh}/c \ll r_{\rm sh}/c$ inside the supershell,
which provides a negligible contribution to the trapping time.} (see
\S~\ref{sec:intro}) through the relation $M_F=t_{\rm trap}/(r_{\rm
sh}/c)$. Indeed, when $v_{\rm sh}\rightarrow 0$ we find that $M_F$
reproduces the value $15(\tau_0/10^{5.5})$ (indicated by {\it
horizontal dotted lines}) that was found by \citet{Adams75} and
\citet{Bo79} reasonably well (keeping in mind that these authors
derived their result for a static, uniform, infinite slab of material,
and assumed different frequency distributions for the emitted
Ly$\alpha$ photons).

\section{Conclusions}
\label{sec:conc}

We have applied an existing Monte-Carlo Ly$\alpha$ radiative transfer code
\citep[described and tested extensively in][]{mc} to the calculation of the
pressure that is exerted by Ly$\alpha$ photons on an optically thick
medium. This code enabled us to perform (the first) direct, accurate calculations of Ly$\alpha$ radiation pressure, which distinguishes this work from previous
discussions on the importance of Ly$\alpha$ radiation pressure in various
astrophysical environments.

We have focused on a range of models which represent galaxies at different
cosmological epochs. In \S~\ref{sec:firststar} we have shown that the
Ly$\alpha$ radiation pressure exerted on the neutral intergalactic medium
(IGM) surrounding minihalos ($M_{\rm tot}\sim 10^6 M_{\odot}$) in which the
first stars form, can exceed gravity by 2--3 orders of magnitude
(Fig~\ref{fig:caseI}), and in principle accelerate the gas in the IGM to
tens of km s$^{-1}$. Thus, Ly$\alpha$ radiation pressure can launch
supersonic winds in the IGM surrounding the first stars. Our analysis did
not allow a detailed study of the consequences of these winds. A
comprehensive study would require numerical simulations that capture the
full IGM density field around minihalos in 3D and track the evolution of
the shocks that may form in the IGM together with the Ly$\alpha$ radiative
transfer. In this paper, we have also shown that Ly$\alpha$ radiation
pressure is important in the neutral IGM that surrounds the HII regions
produced by galaxies with a total halo mass of $M_{\rm tot}=10^8M_{\odot}$
(Fig~\ref{fig:caseII}. These are the lowest mass, and hence the most
abundant, halos in which gas can cool via atomic line excitation. Here,
however, the absolute magnitude of the radiation force is too weak to drive
the gas to supersonic velocities.
 
Finally, we have shown in \S~\ref{sec:wind} that the Ly$\alpha$ radiation
pressure exerted on neutral gas in the interstellar medium (ISM) of a
galaxy can also have strong dynamical consequences. In particular, we have
found that the Ly$\alpha$ radiation force exerted on an expanding HI
supershell can exceed gravity by orders of magnitude
(Fig~\ref{fig:caseIII}), for reasonable assumptions about the
gravitational force. It is therefore possible that Ly$\alpha$ radiation
pressure plays an important role in determining the kinematics of HI
supershells around starburst galaxies. We have demonstrated that the total
Ly$\alpha$ radiation force on a spherical HI supershell can be written as
$F_{\rm rad}=M_F L_{\alpha}/c$, where the 'force-multiplication factor'
$M_F$ relates to the average trapping time of Ly$\alpha$ photons in the
neutral medium. The factor $M_F$ can greatly exceed unity, as illustrated
by Fig~\ref{fig:fscat}. For comparison, the maximum possible radiation
force due to continuum radiation\footnote{
Continuum radiation may exert a force on steady-state outflows
(with a constant mass ejection rate, $\dot{M}$) around late-type stars
that may significantly exceed $\dot{M}\Delta v\gg L_{\rm bol}/c$
\citep[][]{salpeter74,Ivezic95}. This is not because of 'trapping' of
continuum photons, but related to the propagation speed of the photons
and the wind. We similarly expect the radiative force of trapped
Ly$\alpha$ photons in steady-state outflows to potentially exceed
$\dot{M}\Delta v\gg M_FL_{\alpha}/c$ (and as argued in this paper, it
is possible that $M_FL_{\alpha}/c>L_{\rm bol}/c$).
 Note though that these steady-state outflows are clearly different
from those discussed in \S~\ref{sec:wind}, in which a well defined
thin shell HI gas is physically separated from the central Ly$\alpha$
source.}  is $L_{\rm bol}/c$, in which $L_{\rm bol}$ is the
bolometric luminosity of the central galaxy.  For a typical star
forming galaxy, $L_{\alpha}\sim 0.07L_{\rm bol}$ \citep[e.g.][]{PP67},
whereas for a galaxy that contains population III stars,
$L_{\alpha}\sim 0.24L_{\rm bol}$ \citep{S03}. Hence, the Ly$\alpha$
radiation pressure can dominate the maximum possible continuum
radiation pressure if $M_F\gsim 14$ (for a normal stellar population),
a threshold which is easily exceeded at large column densities of
relatively slow-moving HI shells (see Fig~\ref{fig:fscat}).

The possibility that Ly$\alpha$ radiation alone can result in a
radiation force that exceeds $L_{\rm bol}/c$ is
important. \citet{Murray05} have shown that the total momentum carried
by radiation from a star-forming region can exceed the total momentum
deposited by supernova explosions in it, and so galactic outflows
may be driven predominantly by continuum radiation pressure. We have
argued that Ly$\alpha$ radiation pressure may in some cases be even
more important than continuum radiation pressure, and thus provide the
dominant source of pressure on neutral hydrogen in the ISM.

The important implication of our last result is that Ly$\alpha$ radiation
pressure may drive outflows of HI gas in the ISM. Observations of local
starburst galaxies have shown that the presence of outflowing HI gas may be
required to avoid complete destruction of the Ly$\alpha$ radiation by dust
and to allow its escape from the host galaxies
\citep{Kunth98,Hayes08,Ostlin08,Atek08}. At high redshifts, the Ly$\alpha$
emission line of galaxies is often redshifted relative to other nebular
recombination lines (such as H$\alpha$) and metal absorption lines
\citep[e.g.][]{Pettini01,Shapley03}. Furthermore, the spectral shape of the
Ly$\alpha$ emission line is typically asymmetric, with emission extending
well into the red wing of the line \citep[e.g.][]{L95}. Both of these
observations can be explained simultaneously if the observed Ly$\alpha$
photons scatter off neutral hydrogen atoms in an outflowing 'supershell' of
HI gas that surrounds the star forming regions
\citep{L95,T99,Ahn03,Ahn04,Verhamme06,Verhamme08}. The possibility that
Ly$\alpha$ radiation pressure may be important in determining the
properties of expanding supershells is exciting, and is discussed in more
detail in a companion paper (Dijkstra \& Loeb 2008).

{\bf Acknowledgments} 
This work is supported by in part by NASA grant
NNX08AL43G, by FQXi, and by Harvard University funds. We thank Christian Tapken and an anonymous referee for helpful constructive comments.

\appendix

\section{HI Level Populations}
\label{app:n2}
\subsection{The (De)Excitation Rates of the $2p$ Level}

Our calculations assumed that all neutral hydrogen atoms populate
their electronic ground state (i.e. the $1s$ level). Below, we explore
the physical conditions under which this assumption holds.

Processes that populate the $2p$ level include: ({\it i}) collisional
excitation of neutral hydrogen atoms in both the $1s$ and $2s$ states
by electrons and protons, ({\it ii}) recombination into the
$2p$ state following photoionization or collisional ionization, ({\it
iii}) photoexcitation of $2p$ level, ({\it iv}) photoexcitation of
$np$ ($n>3$) levels, followed by a radiative cascade that passes
through the $2p$ state. Processes that {\it de-populate} the $2p$ level
include ({\it v}) collisional de-excitation to by electrons, ({\it vi})
stimulated Ly$\alpha$ emission $2p\rightarrow 1s$ following the
absorption of a Ly$\alpha$ photon, and ({\it vii}) spontaneous
emission of a Ly$\alpha$ photon.

Quantitatively, the rate at which the population in the $2p$ level is
populated is given by

\begin{eqnarray}
\frac{dn_{2p}}{dt}=C_{1s2p}n_en_{1s}+ C_{\rm 2s2p}n_pn_{\rm 2s}+0.68n_en_{{\rm HII}}\alpha_{\rm rec,B}+\\ \nonumber
\sum_{\rm n=2}^{\infty}P_{n}n_{1s}f_{\rm n2}-C_{2p1s}n_en_{2p}-C_{\rm 2p2s}n_pn_{2p}-P_{2}n_{2p}-n_{2p}A_{\rm 21},
\label{eq:rate}
\end{eqnarray} 
where $C_{lu}$'s denote collisional excitation rate coefficients from
level $l$ to $u$ (in cm$^3$ s$^{-1}$), $n_x$ denote number densities
is species 'x' (i.e $n_e$ denotes the electron number density, while
$n_{\rm HII}$ denotes the number density of HII ions), $P_{n}$ denote
photoexcitation rates to the state $np$ (in s$^{-1}$), and $f_{n2}$
denotes the probability that photoexcitation of the $np$ level results
in a radiative cascade that passes through the $2p$ level.

In the reminder of this Appendix, we will estimate the
order-of-magnitude of each term.

\begin{itemize}

\item The Einstein-A coefficient of the Ly$\alpha$ transition is $A_{21}=6.25\times 10^8$ s$^{-1}$. That is, spontaneous emission of Ly$\alpha$ depopulates the $2p$ state at a rate $A_{21}=6.25\times 10^8$ s$^{-1}$.

\item Collisional excitation of neutral hydrogen atoms in $1s$ level
by electrons populates the $2p$ level at a rate
$C_{1s2p}n_e=n_e\left[\frac{8.629\times10^{-6}}{T^{1/2}}\right]
\left[\frac{\Omega(1s,2p)}{g_2}\frac{g_{2p}}{g_{1s}}\right]e^{-\chi/kT}~{\rm
s^{-1}}$ \citep[e.g.][]{Osterbrock89}. Here, $T$ denotes the gas
temperature in K, $n_e$ is the electron density in cm$^{-3}$, $g_{1s}=1$
and $g_{2p}=3$ are the statistical weights of the $1s$ and $2p$ levels,
$\Omega(1s,2p)=0.40-0.50$ \citep[5000 K $<$ T $<$ 2$\times 10^4$ K,][]{Osterbrock89}, and $\chi=10.2$ eV is
the energy difference between the $1s$ and $2p$ levels. For
temperatures $T< 2\times 10^4$ K, we find that $C_{1s2p}n_e<10^{-10}n_e$
s$^{-1}$.

\item The collisional excitation rate of neutral hydrogen atoms in
$2s$ level by protons (which dominate over collisions with electrons
by about a factor of $\sim 10$) populates the $2p$ level at a rate
$C_{2s2p}n_p\sim 2\times 10^{-3} n_p$ s$^{-1}$
\citep[e.g.][]{Osterbrock89}. To assess the term $C_{\rm
2s2p}n_pn_{\rm 2s}$ requires one to compute $n_{2s}$. The fraction of
atoms in the $2s$ state is determined by rates similar to those
mentioned above, except that the $2s$-state is metastable and its
Einstein coefficient is $A_{2s1s}\approx 8$ s$^{-1}$. The $2s$-state may
therefore be overpopulated relative to the $2p$ state by orders of
magnitude \citep[see e.g.][and references therein]{D05}. In close
proximity to a luminous source, the $2s$ level is populated mostly via
transitions of the form $1s\overset{{\rm Ly}\beta}{\rightarrow}3p
\overset{{\rm H}\alpha}{\rightarrow}2s$ \citep{Sethi07,FS}, and
$n_{2s}\sim 0.12 n_{1s} P_3/A_{2s1s}$, where $P_3$ is the rate at which
Ly$\beta$ photons are scattered and the prefactor $0.12$ denotes the
probability that absorption of the Ly$\beta$ is followed by
re-emission of an H$\alpha$ photon \citep[][]{FS}.

\item The recombination rate into the $2p$ level is given by
$0.68\alpha_{\rm rec,B}n_en_p=1.8\times 10^{-13}(T_{\rm
gas}/10^4\hs{\rm K})^{-0.7}n_en_p$ cm$^{3}$ s$^{-1}$
\citep[e.g.][]{Hui97}, where $n_p$ is the proton density in cm$^{-3}$.

\item The rate at which transitions of the form $1s\rightarrow np$
occur by absorbing a photon is given by $P_{n}=4\pi
\int\frac{J(\nu)}{h\nu}\sigma_n(\nu)d\nu$. Assuming for simplicity
that $J(\nu)$ does not vary with frequency, i.e. $J(\nu)=J$, we have
\begin{equation}
P_n=\frac{4\pi J f_n \pi e^2}{h\nu_n m_e c},
\end{equation} 
where $f_n$ denotes the oscillator strength of the transition, $e$
($m_e$) the charge (mass) of the electron, and $h\nu_n$ denotes the
energy difference between the $1s$ and $np$ levels. The oscillator
strength decreases rapidly with increasing $n$ \citep[e.g. chapter
10.5 of][]{RL79}, and in practice we can safely ignore all terms with
$n>2$. We then need not worry about the factors $f_{n2}$ \citep[which
have been computed by][]{PF06,Hirata06}.

The Ly$\alpha$ scattering rate is given by
\begin{equation}
P_2=\frac{M_F L_{\alpha} f_{2} \pi e^2}{4\pi r^2 \Delta \nu h\nu_{\alpha} m_e c},
\end{equation} 
where we replaced $J$ (in erg s$^{-1}$ Hz$^{-1}$ sr$^{-1}$ cm$^{-2}$)
with $J=M_F\frac{L_\alpha}{16\pi^2 r^2 \Delta \nu}$, in which
$L_{\alpha}$ is the Ly$\alpha$ luminosity of the central source (in
erg s$^{-1}$), and $\Delta \nu$ is the frequency range over which
these Ly$\alpha$ photons have been emitted. The factor $M_F$ takes
into account the fact that resonant scattering traps Ly$\alpha$
photons in an optically thick medium (see
\S~\ref{sec:wind}). Substituting fiducial numbers
\begin{equation}
P_2=2 \times 10^2 \hs {\rm
s}^{-1}\hs\Big{(}\frac{L_{\alpha}}{10^{42}\hs{\rm erg}/{\rm
s}}\Big{)}\Big{(}\frac{10^{-3}\nu_{\alpha}}{\Delta
\nu}\Big{)}\Big{(}\frac{{\rm
pc}}{r}\Big{)}^{2}\Big{(}\frac{M_F}{100}\Big{)}.
\label{eq:lya}
\end{equation} 

The rate at which Ly$\beta$ photons scatter can be related to the Ly$\alpha$ luminosity, $L_{\alpha}$, and the equivalent width (EW) of the line, if one writes the specific intensity $J(\nu_{\beta})$ near the Ly$\beta$ resonance in terms of the Ly$\alpha$ luminosity of the central source as $J(\nu_{\beta})=\frac{\lambda_{\beta}}{\nu_{\beta}}\frac{L_{\alpha}}{{\rm EW}}\frac{1}{16\pi^2r^2}$ (note that we assumed that the specific intensity of the continuum remains constant between $\nu_{\alpha}$ and $\nu_{\beta}$). The rate at which Ly$\beta$ photons scatter can then be written as

\begin{equation}
P_3=2 \times 10^{-3} \hs {\rm s}^{-1}\hs\Big{(}\frac{L_{\alpha}}{10^{42}\hs{\rm erg}\hs {\rm s}^{-1}}\Big{)} \Big{(}\frac{{\rm EW}}{200\hs \AA}\Big{)}^{-1}\Big{(}\frac{r}{{\rm pc}}\Big{)}^{-2}.
\label{eq:lyb}
\end{equation} 
Equation~(\ref{eq:lyb}) illustrates that it is very difficult to bring
the ratio $n_{2s}/n_{1s}\sim 0.12 P_3/(A_{\rm 2s1s})$ to unity.

\item Collisional deexcitation rates relate to the collisional
excitation rates via $C_{lu}=C_{ul}\frac{g_u}{g_l}e^{-\chi/kT}$
\citep[e.g.][]{Osterbrock89}. Combined with the formulas given above, it is
straightforward to verify that the collisional de-excitation rates are
subdominant relative to the rate at which spontaneous Ly$\alpha$
emission de-populates the $2p$ level.

\item Lastly, Eq~(\ref{eq:lya}) shows that the stimulated emission rate
is $P_2 \ll A_{21}$.

\end{itemize} 

\subsection{HI Level Populations in this Paper}

The maximum number density of hydrogen nuclei in this paper is
encountered in \S~\ref{sec:wind}, for the expanding shell of HI gas
with $N_{\rm HI}=10^{21}$ cm$^{-2}$ and a thickness $dr=0.01$ kpc, in
which $n_{\rm max}\sim 30$ cm$^{-3}$. At these column densities, the
shell self-shields against ionizing radiation, and is likely mostly
neutral. For simplicity, let us assume that $n_e=n_p=n_{\rm HI}=30$
cm$^{-3}$. Under these conditions:

\begin{itemize}

\item the collisional excitation rate from $1s\rightarrow 2p$ is
$C_{12} < 3\times 10^{-9}$ s$^{-1}$.

\item the collisional excitation rate from $2s\rightarrow 2p$ per atom
in the 1s state- is $C_{2s2p}n_{2s}n_{p}/n_{1s}\sim 5\times
10^{-2}n_{2s}/n_{1s}$ s$^{-1}$$=5\times 10^{-3}P_3/A_{2s1s}$ s$^{-1}$. The maximum
Ly$\alpha$ luminosity considered in this paper is $\sim 10^{43}$ erg
s$^{-1}$. Let us conservatively assume that EW$=20$ \AA (rest-frame),
which corresponds roughly to the detection threshold that exists in
narrow-band surveys \citep[e.g.][]{Shima06}. Using Eq~\ref{eq:lyb}, we
find that $P_{3,{\rm max}} \sim 10^{-5}$ s$^{-1}$ (for $r=0.1$ kpc), 
and therefore that $C_{2s2p}n_{2s}n_{p}/n_{1s}\sim 10^{-8}$ s$^{-1}$.

\item the recombination rate is $5\times 10^{-12}(T_{\rm
gas}/10^4\hs{\rm K})^{-0.7}$ s$^{-1}$.

\item the maximum Ly$\alpha$ scattering rate is (substituting $r=0.1$
kpc, $M_F=100$, $\Delta \nu=0.001\nu_{\alpha}$ into Eq~\ref{eq:lya})
$P_{2,{\rm max}}=0.2$ s$^{-1}$.

\end{itemize} 

By comparing these rates to the rate at which spontaneous emission of
Ly$\alpha$ depopulates the $2p$ state, $A_{21}=6.25\times 10^{8}$
s$^{-1}$, we find that all excitation rates are $\geq 9$ orders of
magnitude smaller than the de-excitation rate for the wind models
discussed in \S~\ref{sec:wind}. In equilibrium, hydrogen atoms in
their electronic ground ($1s$) state are therefore $\geq 9$ orders of
magnitude more abundant than hydrogen atoms in their first excited
($2p$) state. Furthermore, as was mentioned above the ratio of atoms in the 2s and 1s levels is given by $n_{2s}/n_{1s}=0.12 P_{3,{\rm max}}/A_{2s1s}\sim 10^{-7}$. Since the densities, Ly$\alpha$ luminosities, and the Ly$\beta$ scattering rates are lower, the fraction of HI atoms in their first excited states are even smaller in other sections of the paper. In conclusion, for all applications
presented in this paper, no accuracy is lost by assuming that all
hydrogen atoms occupy their electronic ground state.

Finally, in \ref{sec:test} we computed solutions for the radial
dependence of $U(r)$ (Fig~\ref{fig:uden}). The energy density, $U(r)$,
was quoted to depend linearly on the luminosity of the central
source. This is valid unless ({\it i}) the Ly$\alpha$ scattering rate, $P_2> A_{21}=6.25\times 10^8$ s$^{-1}$, or ({\it ii}) the Ly$\beta$ scattering rate exceeds $P_{\rm 3} \gsim 10^2$ s$^{-1}$.  In either case, our
assumption that all hydrogen atoms populate their electronic ground
state breaks down. Condition ({\it i}) translates to $L_{\alpha}\gsim 3\times 10^{48}$ erg s$^{-1}(\Delta \nu/0.001\nu_{\alpha})(r/{\rm pc})^2(100/M_F)$ (Eq.~\ref{eq:lya}), while condition ({\it ii}) translates to $L_{\alpha}\gsim 10^{46}$ erg s$^{-1}({\rm EW}/20\hs{\rm \AA})(r/{\rm
pc})^2$ (Eq.~\ref{eq:lyb}). Substituting $r=0.1$ kpc, $M_F=100$ (Fig~\ref{fig:uden} shows that resonant scattering enhances the energy density by a factor of $\gsim 10$ relative to $L_{\alpha}/4\pi r^2 c$), $\Delta \nu=10^{-3}\nu_{\alpha}$
(thermal broadening alone in $T=10^4$ K gas results in $\Delta v\sim 26$ km s$^{-1}$, which translates to $\Delta \nu=10^{-4}\nu_{\alpha}$), and $EW=20$\AA, condition ({\it i}) translates to $L_{\alpha}\gsim 3\times 10^{52}$ erg s$^{-1}$, while condition ({\it ii}) translates to $L_{\alpha}\gsim 10^{50}$ erg s$^{-1}$
 The more conservative condition ({\it ii}), $L_{\alpha}\gsim 10^{50}$ erg s$^{-1}$, translates to $\dot{N}_{54} \gsim 10^7$, well beyond the regime considered in this paper.
\label{lastpage}

\begin{thebibliography}{xx}
\expandafter\ifx\csname natexlab\endcsname\relax\def\natexlab#1{#1}\fi
\bibitem[Abel et al.(1999)]{Abel99} Abel, T., Norman, M.~L., 
\& Madau, P.\ 1999, \apj, 523, 66 

\bibitem[Abel et al.(2002)]{Abel02} Abel, T., Bryan, G.~L., 
\& Norman, M.~L.\ 2002, Science, 295, 93 

\bibitem[Abel et al.(2007)]{Abel07} Abel, T., Wise, J.~H., 
\& Bryan, G.~L.\ 2007, \apjl, 659, L87 

\bibitem[Adams(1975)]{Adams75} Adams, T.~F.\ 1975, \apj, 201, 
350 

\bibitem[Ahn \& Lee(2002)]{Ahn02} Ahn, S.-H., \& Lee, H.-W.\ 2002, Journal of Korean Astronomical 

\bibitem[Ahn et al.(2003)]{Ahn03} Ahn, S.-H., Lee, H.-W., \& 
Lee, H.~M.\ 2003, \mnras, 340, 863 

\bibitem[Ahn(2004)]{Ahn04} Ahn, S.-H.\ 2004, \apjl, 601, L25 

\bibitem[Atek et al.(2008)]{Atek08} Atek, H., Kunth, D., 
Hayes, M., Ostlin, G., Mas-Hesse, J.~M., 
\& .\ 2008, ArXiv e-prints, 805, arXiv:0805.3501 

\bibitem[Barkana 
\& Loeb(2001)]{BL01} Barkana, R., \& Loeb, A.\ 2001, \physrep, 349, 125 
\bibitem[Barkana(2004)]{Barkana04} Barkana, R.\ 2004, \mnras, 
347, 59 

\bibitem[Bithell(1990)]{Bithell90} Bithell, M.\ 1990, \mnras, 
244, 738 

\bibitem[Bonilha et al.(1979)]{Bo79} Bonilha, J.~R.~M., Ferch, R., Salpeter, E.~E., Slater, G., \& Noerdlinger, P.~D.\ 1979, \apj, 233, 649 

\bibitem[Castor et al.(1975)]{Castor75} Castor, J.~I., Abbott, D.~C., \& Klein, R.~I., 1975, \apj, 195, 157 

\bibitem[Chandrasekhar(1945)]{Chandra45} Chandrasekhar, S.\ 1945, 
\apj, 102, 402 

\bibitem[Chen et al.(2007)]{Chen07} Chen, H.-W., Prochaska, 
J.~X., \& Gnedin, N.~Y.\ 2007, \apjl, 667, L125 

\bibitem[Cox(1985)]{Cox85} Cox, D.~P.\ 1985, \apj, 288, 465
 
\bibitem[Dennison et al.(2005)]{D05} Dennison, B., Turner, 
B.~E., \& Minter, A.~H.\ 2005, \apj, 633, 309 

\bibitem[Dijkstra et al.(2006)]{mc} Dijkstra, M., Haiman, 
Z., \& Spaans, M.\ 2006, \apj, 649, 14 

\bibitem[Dijkstra et al.(2007)]{igm} Dijkstra, M., Lidz, 
A., \& Wyithe, J.~S.~B.\ 2007, \mnras, 377, 1175 

\bibitem[Dijkstra \& Loeb(2008)]{dl} Dijkstra, M., \& Loeb, A.
2008, submitted to \mnras

\bibitem[Dijkstra et al.(2008)]{FS} Dijkstra, M., et al.
2008, accepted to \mnras

\bibitem[Elitzur \& Ferland(1986)]{EF86} Elitzur, M., \& Ferland, G.~J.\ 1986, \apj, 305, 35 

\bibitem[Furlanetto et al.(2004)]{F04} Furlanetto, S.~R., 
Zaldarriaga, M., \& Hernquist, L.\ 2004, \apj, 613, 1 

\bibitem[Gnedin et al.(2008)]{Gnedin08} Gnedin, N.~Y., Kravtsov, A.~V., \& Chen, H.-W.\ 2008, \apj, 672, 765 

\bibitem[Haehnelt(1995)]{Haenelt95} Haehnelt, M.~G.\ 1995, 
\mnras, 273, 249 

\bibitem[Haiman et al.(1996)]{Haiman96} Haiman, Z., Thoul, 
A.~A., \& Loeb, A.\ 1996, \apj, 464, 523 

\bibitem[Hayes et al.(2008)]{Hayes08} Hayes, M., Ostlin, G., 
Mas-Hesse, J.~M., \& Kunth, D.\ 2008, ArXiv e-prints, 803, arXiv:0803.1176 

\bibitem[Heger \& Woosley(2002)]{HW02} Heger, A., \& Woosley, S.~E.\ 2002, \apj, 567, 532 

\bibitem[Heiles(1984)]{Heiles84} Heiles, C.\ 1984, \apjs, 55, 
585 

\bibitem[Hirata(2006)]{Hirata06} Hirata, C.~M.\ 2006, \mnras, 
367, 259 

\bibitem[Hui \& Gnedin(1997)]{Hui97} Hui, L., \& Gnedin, N.~Y.\ 1997, \mnras, 292, 27 

\bibitem[Ivezic \& Elitzur(1995)]{Ivezic95} Ivezic, Z., \& Elitzur, M.\ 1995, \apj, 445, 415

\bibitem[Kitayama et al.(2004)]{Kita04} Kitayama, T., Yoshida, 
N., Susa, H., \& Umemura, M.\ 2004, \apj, 613, 631 

\bibitem[Komatsu et al.(2008)]{Komatsu08} Komatsu, E., et al.\ 
2008, ArXiv e-prints, 803, arXiv:0803.0547 

\bibitem[Kunth et al.(1998)]{Kunth98} Kunth, D., Mas-Hesse, J.~M.,
Terlevich, E., Terlevich, R., Lequeux, J., \& Fall, S.~M.\ 1998, \aap, 334,
11

\bibitem[Lee 
\& Ahn(1998)]{Lee98} Lee, H.-W., \& Ahn, S.-H.\ 1998, \apjl, 504, L61 

\bibitem[Lequeux et al.(1995)]{L95} Lequeux, J., Kunth, D., Mas-Hesse,
J.~M., \& Sargent, W.~L.~W.\ 1995, \aap, 301, 18

\bibitem[Loeb \& Rybicki(1999)]{LR99} Loeb, A., \& Rybicki, G.~B.\ 1999,
\apj, 524, 527 (LR99)

\bibitem[Loeb(2001)]{Loeb01} Loeb, A.\ 2001, \apjl, 555, L1 

\bibitem[Lucy(1999)]{Lucy99} Lucy, L.~B.\ 1999, \aap, 344, 282 

\bibitem[Lucy(2007)]{Lucy07} Lucy, L.~B.\ 2007, \aap, 468, 649 

\bibitem[Maller \& Bullock(2004)]{Maller04} Maller, A.~H., \& Bullock,
J.~S.\ 2004, \mnras, 355, 694

\bibitem[McKee \& Tan(2008)]{McKee07} McKee, C.~F., \& Tan, J.~C.\ 2008, \apj, 681, 771 

\bibitem[McClure-Griffiths et al.(2002)]{M02} 
McClure-Griffiths, N.~M., Dickey, J.~M., Gaensler, B.~M., 
\& Green, A.~J.\ 2002, \apj, 578, 176 

\bibitem[McQuinn et al.(2007)]{McQuinn07} McQuinn, M., Lidz, A., Zahn, O., Dutta, S., Hernquist, L., 
\& Zaldarriaga, M.\ 2007, \mnras, 377, 1043 

\bibitem[Murray et al.(2005)]{Murray05} Murray, N., Quataert, 
E., \& Thompson, T.~A.\ 2005, \apj, 618, 569 

\bibitem[Oh \& Haiman(2002)]{Oh02} Oh, S.~P., \& Haiman, Z.\ 2002, \apj, 569, 558

\bibitem[Osterbrock(1989)]{Osterbrock89} Osterbrock, D.~E.\ 1989, {\it Astrophysics of gaseous nebulae and active galactic nuclei}, University of Minnesota, et al.~Mill Valley, CA, University Science Books.  

\bibitem[Ostlin et al.(2008)]{Ostlin08} Ostlin, G., Hayes, M., 
Kunth, D., Mas-Hesse, J.~M., Leitherer, C., Petrosian, A., 
\& Atek, H.\ 2008, ArXiv e-prints, 803, arXiv:0803.1174 

\bibitem[Ouchi et al.(2008)]{Ouchi08} Ouchi, M., et al.\ 2008, 
\apjs, 176, 301 

\bibitem[Partridge \& Peebles(1967)]{PP67} Partridge, R.~B., \& Peebles, P.~J.~E.\ 1967, \apj, 147, 868 

\bibitem[Pettini et al.(2001)]{Pettini01} Pettini, M., Shapley, 
A.~E., Steidel, C.~C., Cuby, J.-G., Dickinson, M., Moorwood, A.~F.~M., 
Adelberger, K.~L., \& Giavalisco, M.\ 2001, \apj, 554, 981 

\bibitem[Pritchard \& Furlanetto(2006)]{PF06} Pritchard, J.~R., \& Furlanetto, S.~R.\ 2006, \mnras, 367, 1057 

\bibitem[Pritchard 
\& Loeb(2008)]{PL08} Pritchard, J.~R., \& Loeb, A.\ 2008, ArXiv e-prints, 802, arXiv:0802.2102 

\bibitem[Rybicki \& Lightman(1979)]{RL79} Rybicki, G.~B., \& Lightman, A.~P.\ 1979, New York, Wiley-Interscience, 1979.~393 p.,  

\bibitem[Ryder et al.(1995)]{Ryder95} Ryder, S.~D., 
Staveley-Smith, L., Malin, D., \& Walsh, W.\ 1995, \aj, 109, 1592 

\bibitem[Salpeter(1974)]{salpeter74} Salpeter, E.~E.\ 1974, \apj, 193, 585

\bibitem[Schaerer(2002)]{S02} Schaerer, D.\ 2002, \aap, 382, 28 

\bibitem[Schaerer(2003)]{S03} Schaerer, D.\ 2003, \aap, 397, 527 

\bibitem[Sethi et al.(2007)]{Sethi07} Sethi, S.~K., 
Subrahmanyan, R., \& Roshi, D.~A.\ 2007, \apj, 664, 1 

\bibitem[Shapley et al.(2003)]{Shapley03} Shapley, A.~E., 
Steidel, C.~C., Pettini, M., \& Adelberger, K.~L.\ 2003, \apj, 588, 65
 
\bibitem[Shimasaku et al.(2006)]{Shima06} Shimasaku, K., et 
al.\ 2006, \pasj, 58, 313 


\bibitem[Tapken et al.(2007)]{Tapken07} Tapken, C., Appenzeller, I., Noll,
S., Richling, S., Heidt, J., Meink{\"o}hn, E., \& Mehlert, D.\ 2007, \aap,
467, 63

\bibitem[Tenorio-Tagle \& Bodenheimer(1988)]{T88} Tenorio-Tagle, G., \&
Bodenheimer, P.\ 1988, \araa, 26, 145

\bibitem[Tenorio-Tagle et al.(1999)]{T99} Tenorio-Tagle, 
G., Silich, S.~A., Kunth, D., Terlevich, E., 
\& Terlevich, R.\ 1999, \mnras, 309, 332 

\bibitem[Verhamme et al.(2006)]{Verhamme06} Verhamme, A., 
Schaerer, D., \& Maselli, A.\ 2006, \aap, 460, 397 

\bibitem[Verhamme et al.(2008)]{Verhamme08} Verhamme, A., Schaerer, D.,
Atek, H., \& Tapken, C.\ 2008, ArXiv e-prints, 805, arXiv:0805.3601

\bibitem[Wyithe \& Loeb(2006)]{Wyithe06} Wyithe, J.~S.~B., \& Loeb, A.\ 2006, \nat, 441, 322 

\bibitem[Yoshida et al.(2006)]{Yoshida06} Yoshida, N., Omukai, 
K., Hernquist, L., \& Abel, T.\ 2006, \apj, 652, 6 

\end{thebibliography}
\end{document}